\documentstyle[12pt,aaspp4,flushrt]{article}
\def\etal{{\it et~al.}}
\def\cgs{erg~s$^{-1}$cm$^{-2}$sr$^{-1}$}

\begin{document}

\title
{[C {\small II}] Emission From NGC 4038/39 (The ``Antennae'')}

\author
{T.~Nikola\altaffilmark{1}, R.~Genzel, F.~Herrmann,
S.C.~Madden\altaffilmark{2}, A.~Poglitsch}
\affil
{Max-Planck-Institut f{\"u}r extraterrestrische Physik, Garching, Germany}
\and 
\author
{N.~Geis\altaffilmark{3}, C.H.~Townes}
\affil 
{Department of Physics, University of California, Berkeley, CA 94720}
\and
\author
{G.J.~Stacey}
\affil
{Department of Astronomy, Cornell University, Ithaca, NY 14853}
\altaffiltext{1}{present address: University of New South Wales, Sydney,
Australia}
\altaffiltext{2}{present address: CE-Saclay, Service d'Astrophysique, 91191
Gif-sur-Yvette, France}
\altaffiltext{3}{present address: Max-Planck-Institut f{\"u}r 
extraterrestrische Physik, Garching, Germany}

\begin{abstract}

We present observations of NGC 4038/39 in the [\ion{C}{2}] 158 \micron \/
fine structure line taken with the MPE/UCB Far-infrared Imaging Fabry-Perot 
Interferometer (FIFI) on the KAO.
A fully sampled map of the galaxy pair (without the tidal tails) at
55\arcsec \ resolution has been obtained.
The [\ion{C}{2}] emission line is detected from the entire galaxy pair and
peaks at the interaction zone.
The total [\ion{C}{2}] luminosity of the Antennae is $L_{\rm [C II]} =
3.7 \times 10^{8} L_{\sun}$, which is about 1\% of the far-infrared
luminosity observed with IRAS.
The main part of the [\ion{C}{2}] emission probably arises from
photodissociation regions (PDRs), and a minor fraction may be emitted from
\ion{H}{2} regions.
A small part of the [\ion{C}{2}] emission comes from standard cold neutral
medium (CNM); however, for high temperature ($T \sim 100$~K) and high
density ($n_{\rm H} \sim  200$~cm$^{-3}$) about one
third of the observed [\ion{C}{2}] emission may originate from CNM.
From PDR models we derive densities of the order of $\sim 10^{5}$~cm$^{-3}$
and far-UV (FUV) intensities of $460\chi_{\circ}$, $500\chi_{\circ}$, and
$240\chi_{\circ}$ for the PDRs in the interaction zone, NGC~4038, and
NGC~4039, respectively.
However, PDRs with densities of the order of $\sim 10^{2}$~cm$^{-3}$ and FUV
intensities of the order of $\sim 100\chi_{\circ}$ could
also explain the observed [\ion{C}{2}] emission.
The minimum masses in the [\ion{C}{2}] emitting regions in the interaction
zone and the nuclei are a few $\times 10^{7}~M_{\odot}$.
A comparison with single dish CO observations of the Antennae shows a
[\ion{C}{2}] to CO intensity ratio at the interaction zone a 
factor of 2.6 lower than usually observed in starburst galaxies, but still
a factor of about 1.3 to 1.4 higher than that at the nuclei of NGC 4038/39.
Therefore, no global starburst is taking place in the Antennae.
[\ion{C}{2}] emission arising partly from confined starburst regions and 
partly from surrounding quiescent clouds could explain the observed
[\ion{C}{2}] radiation at the interaction zone and the nuclei, though the
star formation activity toward the nuclei is lower.
Accordingly there are small confined regions with high star formation
activity in the interaction zone and with a lower star formation activity
in the nuclei.
This supports the high density and high FUV intensity for the PDRs in the
interaction zone and the nuclei.

\end{abstract}


\section{Introduction}

The galaxy pair NGC 4038/39 (Arp 244) is
an interacting system in an early stage of merging at a distance of about 
21 Mpc from our own galaxy.
On long-exposed images in the optical (e.g. Arp 1966) the interaction
is clearly visible because of the tails (``Antennae'') emerging from two 
uniformly luminous, partly overlapping ovals and because of the dwarf
galaxy that appears to have formed at the tip of the southern tail through
the interaction (Zwicky 1956, Schweizer 1978, Mirabel, Dottori, \& Lutz 
1992). 
The tails contain about 70~\% of the total amount of \ion{H}{1} in the
system (van der Hulst 1979).
Short-exposed images (Laustsen, Madsen, \& West 1987) reveal hints of the 
interaction on somewhat smaller scales; the distorted arrangement of 
H$\alpha$ knots and the velocity distribution of the individual knots 
lead Rubin \etal \ (1970) to conclude there is an interaction of two 
rotating galaxies.
Computer simulations carried out in the classical paper of Toomre \& Toomre
(1972) and later by Barnes (1988) can account for the present morphological
appearance of the system quite well by assuming an interaction of two
rotating spiral galaxies.

Spectra of the nuclei of both galaxies taken in the optical range do not 
look like pure starburst spectra but consist of a composition of early-type 
stars and late giants (Keel \etal \ 1985).
The detection of bright near-infrared peaks at the nuclei lead Bushouse \&
Werner (1990) to the same result.
Their images of the Antennae in J- and R-band and in H$\alpha$ show also the 
same pattern of bright knots in the surroundings of NGC 4038 and in the 
bridge connecting both galaxies. 
The Antennae system as a whole shows a relatively low star forming
efficiency according to the ratio $L_{\rm IR} /M({\rm H}_2) \approx 8.55$
measured by Young \etal \ (1986); it is only a factor of 3 higher than in
the Milky Way.
The ratio determined by Sanders \& Mirabel (1985) is almost twice as high.
However, they observed a smaller region in CO and therefore probably
underestimated the molecular mass.
Comparison of the Antennae with the sample of interacting and isolated
galaxies of Young \etal \ (1986) shows NGC 4038/39 to have characteristics
more like isolated galaxies.

Measurements in the radio continuum at 1.5~GHz and 4.9~GHz of 
NGC~4038/39 (Hummel \& van der Hulst
1986) reveal a number of discrete knots which coincide in general with 
H$\alpha$ knots, and an underlying diffuse component.
This diffuse component has a steep spectral index on average which
indicates non-thermal emission, and the peak of the diffuse radio emission
is at the dust patch near the overlapping region.
The discrete radio knots account for roughly 35~\% of the total radio
emission and have a spectral index of $\alpha \approx -0.5$ on average,
probably due to a thermal contribution (Hummel \& van der Hulst 1986).

Interferometric observations of the Antennae in CO ($1 \to 0$) by Stanford 
\etal \ (1990) show three main concentrations of CO emission.
Two are associated with the nuclei and the third with the interaction zone.
The overlap region is the strongest CO source and contains $\approx 
10^{9}~M_{\sun}$ of gas, roughly as much H$_{2}$ as both nuclei together.
Based on 10$\mu$m and H$\alpha$ data, the authors have calculated a star 
forming rate of $5 M_{\sun}$~yr$^{-1}$, and consequently 
the life time of the molecular gas is $2 \times 10^{8}$~yr.
Single dish observations in CO were made at the interaction zone and both 
nuclei of the Antennae by Aalto \etal \ (1995).
The ratio of the emission lines of $^{12}$CO and $^{13}$CO measured in the 
nuclei and the 
overlapping region of the Antennae is similar to that found in the
central regions of ``normal'' starburst galaxies (Aalto \etal \ 1995).

An excellent tracer of star formation activity in galaxies is the strong 
[\ion{C}{2}] 158~$\mu$m $^{2}$P$_{3/2}\to \, ^{2}$P$_{1/2}$ fine structure
line which arises mainly from photodissociation regions (PDRs) created by
far-ultraviolet photons from hot young stars impinging on nearby dense
interstellar clouds (Crawford \etal \ 1985, Stacey \etal \ 1991).
In combination with CO and FIR observations, the [\ion{C}{2}] emission can
be used with PDR models (Tielens \& Hollenbach 1985, Wolfire, Hollenbach, 
\& Tielens 1989, Wolfire, Tielens, \& Hollenbach 1990) to derive densities 
and far-UV intensities and estimates of the star formation activity.
Extragalactic surveys of [\ion{C}{2}] emission (Crawford \etal \ 1985, 
Stacey \etal 1991) concentrated mainly on nuclei while more recent 
observations have imaged individual galaxies to study the distribution of
[\ion{C}{2}] emission on large scales.
[\ion{C}{2}] images of M83 (Geis \etal \ 1998) and NGC~6946 (Madden \etal \
1993) demonstrate that the emission is extended at least over the full 
optical extent and often follows the distribution of the FIR and CO within
the disk of the galaxies.
In the case of NGC~6946 [\ion{C}{2}] emission beyond the optical extent of
the galaxy has been found.
This [\ion{C}{2}] emission has been attributed to diffuse gas and not to
PDRs.
Therefore we also estimate the possible contribution of [\ion{C}{2}] 
emission from neutral atomic gas and from ionized gas in NGC~4038/39.
Because of its relative proximity, NGC~4038/39 is a unique source
for carrying out spatially resolved measurements of the [\ion{C}{2}] line
in an interacting system.
We present the results of our imaging spectroscopy study of the [\ion{C}{2}]
line in NGC~4038/39 and compare them with observations obtained with ISO.



\section{Observations}

The observations of the Antennae in the [\ion{C}{2}] 
$^{2}$P$_{3/2}\to \, ^{2}$P$_{1/2}$ fine structure line at 157.7409~\micron
\ have been carried out during four individual flights in 1992 with the
Kuiper Airborne Observatory (KAO) from Christchurch, New Zealand, using 
the MPE/UCB Far-infrared Imaging Fabry-Perot Interferometer (FIFI;
Poglitsch \etal \ 1991, Stacey \etal \ 1992).
At 158~\micron \ the FWHM of the beam is about 55\arcsec.
The detector array consists of 5~$\times$~5 pixels of size
40\arcsec~$\times$~40\arcsec \ per pixel on the sky.
We observed three array positions to cover an area of
300\arcsec~$\times$~300\arcsec \ including both galaxies and the
interaction zone and to fully sample an area of
150\arcsec~$\times$~150\arcsec \ around the interaction zone.
Due to the large velocity dispersion in one array setting, as determined
from the emission of the H$\alpha$ knots (Rubin, Ford, \& D'Odorico 1970)
and from the CO emission (Stanford \etal \ 1990),
we chose a spectral resolution of 144~km~s$^{-1}$ (FWHM) and a scan width of
380~km~s$^{-1}$.
The scan center was set at 1600~km~s$^{-1}$.
The wavelength calibration was implemented by means of the H$_{2}$S
absorption line at 157.7726~\micron.
Flat fielding of the detector was carried out with two internal black bodies.
The secondary mirror of the telescope was chopped approximately 4\arcmin \
in roughly the east-west direction, and the telescope was nodded to 
compensate for the beam offset.
For absolute intensity calibration we observed Jupiter at 158~\micron \ and
assumed a temperature of 128 K at this wavelength (Hildebrand \etal \ 1985),
an equatorial diameter of 43.22\arcsec , and a pole diameter of
40.42\arcsec \ (The Astronomical Almanac 1992) for Jupiter.
The accuracy of the intensity calibration is estimated to be about 30~\% and 
the absolute pointing positions are uncertain by about 15\arcsec.

\section{Data Reduction}

The Fabry-Perot interferometers in FIFI are adjusted to the appropriate
velocity range and spectral resolution for every object.
Further the Fabry-Perot interferometers are put to a 'park position' when 
the instrument is not in use.
Therefore the Fabry-Perot interferometers are adjusted newly for every
observing flight and for every object and there may occure slight shifts
($\approx $ a few km/s) of the velocity center of the spectral scans between 
individual observations.
To combine the different pointings of the observation of the Antennae 
obtained in different observing flights
we used the velocity range contained in all observations 
(1420 -- 1740~km~s$^{-1}$). 

To combine the different array positions, to account for the statistical 
errors in the data, and to regrid the data points on a regular grid, we
constructed a program using the method of maximum entropy. 
The program follows mainly the procedure of Gull \& Daniel (1978) as 
described in Skilling \& Bryan (1984).
It determines the maximum entropy of an image consisting of model data
points on a pre-set regular grid on condition that the model data, after 
convolving with the beam, fits the observed data according to a 
chi-square fit.
For large N, the program stops if $\chi^{2}$ is approximately equal to 
$N + 3.29 \sqrt{N}$ where $N$ is the number of observed data points 
(Skilling \& Bryan 1984).
This criterion corresponds to a 99~\% confidence level for the chi-square
fit.
The observed data points were deconvolved and regridded with this program 
onto a 10\arcsec \ regular grid and smoothed with a 55\arcsec \ Gaussian 
beam. 

Taking the peculiar velocity structure of the Antennae (as seen, for 
example, in H$\alpha$ (Amram \etal \ 1992)) into account and using spectral
separation we can distinguish different components of the system that are 
not spatially resolved.
Therefore we subdivided the total velocity range into two velocity ranges
($v =$ 1420 -- 1554~km~s$^{-1}$ and $v =$ 1550 -- 1740~km~s$^{-1}$) in
order to distinguish between the interaction zone (at roughly $v =$ 1450
-- 1590~km~s$^{-1}$) and the nuclei (at $v =$ 1600~km~s$^{-1}$ and $v =$
1630~km~s$^{-1}$ for NGC 4039 (the southern galaxy) and NGC 4038 (the 
northern galaxy), respectively) (Amram \etal \ 1992, Rubin \etal \ 1970, 
Stanford \etal \ 1990).
The cut in the velocity range is somewhat arbitrary since on the one hand 
the velocity gradient between the interaction zone and NGC 4039 is large, 
and on the other hand no clear spatial transition in the velocity occurs 
between NGC 4038 and the interaction zone.

To obtain the [\ion{C}{2}] integrated intensity maps in different velocity
ranges we subdivided the raw data into two adjacent velocity bins and then 
applied the maximum entropy program for each velocity range separately. 
Since the upper and lower velocity ranges are derived only from part of the
spectrum they have slighty different $\sigma$ values.
The $\sigma$ values for the total, the upper, and the lower velocity ranges 
are $1.04 \times 10^{-5}$~\cgs, $0.77 \times 10^{-5}$~\cgs, and
$0.71 \times 10^{-5}$~\cgs, respectively.

\section{Results}

\subsection{Spatial Distribution of the [C II] Emission}
\label{se:spa}

We superimposed maps of the total [\ion{C}{2}]~158~\micron \ line emission 
and the emission in the two seperate velocity ranges on an optical image
(Laustsen, Madsen, \& West 1987) of the Antennae 
(Fig.1a~--~c). 
For a better comparison we used the same contour levels in each velocity
range.
The contour levels are in steps of 1 $\sigma$ deduced from the data of the 
total velocity range.
C$^{+}$ is seen over the full extent of the merging system 
(Fig.1a).
From the line intensity integrated over the total velocity range one can
infer the origin of the bulk of the [\ion{C}{2}] emission.
In all of the three images the peak of the [\ion{C}{2}] emission is situated
at the interaction zone with 
slight differences in the location of the [\ion{C}{2}]
peak in each velocity range.
The distribution of the [\ion{C}{2}] emission in the total velocity range 
(Fig.1a) and the upper velocity range (corresponding to the 
nuclei) (Fig.1b) are very similar and agree well with the
overall shape of the
galaxy pair as seen in short-exposed optical images, if we take the 
different beam sizes into account. 
The emission is elongated along the bridge between the two galaxies, 
including the interaction zone, NGC~4039, and extends even further south.
It is also extended in the north-west showing the western loop of
[\ion{H}{2}] regions around NGC~4038.
However, for the upper velocity range one would have expected a stronger 
concentration of the [\ion{C}{2}] emission at the nuclei and less strong 
at the interaction zone.
Since the cut between the velocity ranges was made at 1550~km~s$^{-1}$ 
and the velocity range of the gas in the interaction zone is 
1450--1590~km~s$^{-1}$ (Amram \etal \ 1992; Rubin,
Ford, \& D'Odorico 1970; Stanford \etal \ 1990) it may be possible that most
of the [\ion{C}{2}] emission is coming from gas with velocities higher than 
1550 ~km~s$^{-1}$.
This would also explain the relatively weak [\ion{C}{2}] emission in the
lower velocity range.
The [\ion{C}{2}] emission in the lower velocity range is less extended than
in the other velocity ranges and more concentrated toward the interaction 
zone as expected from that velocity.
The [\ion{C}{2}] peak coincides also with the peak of the CO emission of the
Antennae (Stanford \etal \ 1990) (Fig.2).
Within the interaction zone the southern clump is the strongest CO emitting
region.

\subsection{Estimate of the Parameters of the [C II] Emitting Region}
\label{se:phys}

The results of the measurement of the [\ion{C}{2}]~158~\micron \ emission
in the different velocity ranges for different positions are given in 
Table 1.
The total [\ion{C}{2}] luminosity from the region enclosed by the first 
contour line in Fig.1a is $L_{\rm [C II]} =
3.7 \times 10^{8} L_{\sun}$ for the total velocity range.
In comparison with the [\ion{C}{2}] ISO-LWS observations of the Antennae
(Fischer \etal \ 1996), we determined a line flux of $\sim 9 \times 
10^{-19}$~W~cm$^{-2}$ within the area of the larger ISO beam. 
This is about 2.4 times higher than the line flux measured with ISO.
By weighting the data measured with FIFI within the ISO beam with a 
gaussion of $80\arcsec$ FWHM we obtain a line flux of $6.5 \times 
10^{-19}$~W~cm$^{-2}$ which agrees within the statistical errors and the
calibration errors with the value measured with ISO.
A possible explanation for this discrepancy would be if the [\ion{C}{2}] 
emission originates from small clumped regions spread over the 
interaction zone.
Thus the emission of clumps at the edge of the ISO beam are only accounted
for with the weight of the beam.


The [\ion{C}{2}]~158~\micron \ fine structure line may arise 
from at least three components of the interstellar medium: photodissociation 
regions (PDRs), atomic gas clouds, and diffuse \ion{H}{2} regions.
In starburst galaxies and star forming regions the contribution of the 
[\ion{C}{2}] fine structure line from PDRs is much stronger than from 
other components of the interstellar medium which can therefore be neglected.
However, it is important to investigate the origin of the [\ion{C}{2}] 
emission from the NGC 4038/39 system.
We follow a procedure originally presented in Madden \etal \ (1993), to 
investigate the various origins of [\ion{C}{2}] emission.

\subsubsection{Estimate of the FIR continuum}
\label{se:FIR}

No FIR continuum map with a resolution comparable to the [\ion{C}{2}] data 
exists for the Antennae.
The only FIR continuum data available to compare with the [\ion{C}{2}] 
emission is from IRAS observations.
According to the formula given in Lonsdale \etal \ (1985) 
\begin{displaymath}
 {\rm FIR} = 1.26 \times (2.58 \times 10^{-14} \times 
 S_{60} + 1 \times 10^{-14} \times S_{100}) \qquad \lbrack 
 {\rm W~m}^{-2} \rbrack
\end{displaymath}
we calculated the FIR luminosity from the IRAS flux densities at 60~$\mu $m 
and 100~$\mu $m for the entire Antennae system.
With $S_{60} = 48.68$~Jy and $S_{100} = 82.04$~Jy (Surace \etal \ 1993), 
the FIR luminosity for the total system is $L_{\rm FIR} = 3.6 \times 10^{10} 
L_{\sun}$, with an assumed distance of 21~Mpc.
Consequently the total [\ion{C}{2}] luminosity is about 1\% of the FIR 
luminosity.
An attempt to obtain sufficient spatial resolution to resolve the nuclei 
from the IRAS data using HIRES and ADDSCAN was unsuccessful (Surace \etal \ 
1993).
Therefore, to get an estimate of the FIR continuum separately for the 
interaction zone and the nuclei we use the FIR/radio correlation.
However, in general the FIR/radio correlation is only valid for 
late-type galaxies.
Despite this we use this correlation because the optical spectra of the
nuclei indicate a mixture of \ion{H}{2} regions with an old population and
the interaction zone shows also a strong radio continuum (Hummel \& van der 
Hulst 1986) in combination with the young star forming region.
The ISOCAM observations of the 6.7 and 15~\micron \ continuum reveal very 
clumpy emission dominated by bright knots, and from the good correlation
of the 15~\micron \ continuum intensity with the [\ion{Ne}{3}]/[\ion{Ne}{2}]
ratio, Vigroux \etal \ (1996) deduce that the 15~\micron \ continuum is due 
to thermal emission from hot dust heated by the absorption of the ionizing 
photons emitted by young stars.
On the kpc scale of our beam, the extension of the FIR continuum originating 
from thermal emission from moderately warm dust, and the 15~\micron \ 
continuum created from these young star forming regions may be similar.
Therefore, a strong but very confined, recent starburst as it is seen at the 
interaction zone with ISOCAM probably does not distort the FIR/radio 
correlation very much on our kpc scale.

For the whole Antennae the FIR luminosity and the total flux density of
the 20~cm radio continuum, $S_{1.5 {\rm GHz}} = 486 \pm 20$~mJy 
(Hummel \& van der Hulst 1986), fulfil the FIR/radio correlation: 
\begin{displaymath}
 \log \frac{P_{1.5 {\rm GHz}}}{{\rm W Hz}^{-1}} = a \times \log 
 \frac{L_{\rm FIR}}{\rm W} + b
\end{displaymath}
with $a = 1.30 \pm 0.03$, $b = -25.82$ and $P_{1.5 {\rm GHz}} = 
S_{1.5 {\rm GHz}} \times 4 \pi R^{2}$ (Xu \etal \ 1994). 
Under the assumption that the FIR/radio correlation is also valid at the 
scale of our beam size we estimate the FIR continuum at the interaction zone 
and the nuclei by distributing the FIR continuum the same way as the low
spatial resolution 20~cm radio continuum (Hummel \& van der Hulst 1986)
(Fig.~\ref{fig:C II20}).
For that we convolved the 20~cm map to a spatial resolution of 55\arcsec \
and scaled it such that the total flux in the map matches the IRAS 
observation.
We obtain a FIR continuum luminosity within our beam of $L_{\rm FIR} = 2.0 
\times 10^{10} L_{\sun}$ at the [\ion{C}{2}] peak, $L_{\rm FIR} = 1.4 \times 
10^{10} L_{\sun}$ at NGC~4038, and $L_{\rm FIR} = 8 \times 10^{9} L_{\sun}$ 
at NGC~4039.


\subsubsection{[C II] Emission From PDRs}

For the treatment of PDRs we used the model described by Stacey \etal \ 
(1991, and references therein). 

Photodissociation regions are the interfaces between \ion{H}{2} regions 
and molecular clouds where photons with energies less than 13.6~eV escape
from the \ion{H}{2} regions, dissociate molecules and ionize elements with
dissociation or ionization energies lower than the Lyman limit (13.6~eV).
Carbon is the most abundant element with an ionization energy (11.3~eV) less 
than the 13.6~eV.
Ionized carbon is then excited by collisions with electrons and/or atomic 
and molecular hydrogen.
The gas in this region is heated mainly by photoelectric emission from
grains illuminated by far-UV radiation. 
However, most of the far-UV flux which is absorbed by grains is converted 
into FIR radiation.

Assuming high temperature ($T \gg 91$~K), high density ($n \gg n_{\rm crit}
= 3.5 \times 10^{3} {\rm cm}^{-3}$), optically thin [\ion{C}{2}] emission, 
and a beam filling factor of unity we can estimate a lower limit of the 
C$^{+}$ column density and hydrogen mass (Crawford \etal \ 1985).
Using a solar fractional carbon abundance of [C]/[H]~$\sim 3 \times 10^{-4}$ 
and assuming that all the carbon is in form of C$^{+}$ we derive a lower 
limit of the hydrogen column density.
The result of this estimate is given in Table 2.
For each of the positions we took the integrated [\ion{C}{2}] intensity
in the total velocity range.
A comparison of the estimated masses of the PDRs with the molecular masses
given by Stanford \etal \ (1990) show that the ratio of the mass of the PDR
to the molecular mass is 6\% for the whole Antennae, 6\% for the interaction
zone, 4\% for NGC~4038, and 15\% for NGC~4039.
The high ratio for NGC~4039 is probably due to the fact that the beam of the 
[\ion{C}{2}] observation also includes part of the interaction zone.
Except for the high ratio in NGC~4039 the PDR to molecular gas mass ratio lies
within the expected range of 1--10\% (Stacey \etal \ 1991).

To compare the [\ion{C}{2}] integrated intensity with the CO ($1\to 0$) 
intensity we use the CO-data from Aalto \etal \ (1995) since their 
observations were made with a single dish and with a similar beamsize, 
43\arcsec \ FWHM. 
They observed three different positions: the interaction zone, NGC 4039, and 
NGC 4038.
For the comparison, we use the [\ion{C}{2}] emission in the total velocity 
range since the measurement of Aalto \etal \ (1995) was also obtained in a 
wide velocity range ($\sim 1300$~km~s$^{-1}$).

In Figure 4, we plotted $Y_{\rm [C II]} = I_{\rm [C II]} / 
\chi_{\rm FIR}$ versus $Y_{\rm CO} = I_{\rm CO} / \chi_{\rm FIR}$.
The advantage of taking these ratios is that they allow one to determine the 
gas density, $n_{\rm H}$, and UV-field, $\chi_{\rm UV}$, independently of 
the beam filling factor if it is assumed that the [\ion{C}{2}], the CO, and 
the FIR emission arise from the same regions.
The theoretically calculated ratios for a set of constant densities
$n_{\rm H}$ (solid lines) and a set of constant UV intensities
$\chi_{\rm UV}$ (6~eV $\le h\nu \le $13.6~eV) (thick dashed lines) are also 
shown in Figure~\ref{fig:pdr}
(Wolfire, Hollenbach , \& Tielens (1989), Stacey \etal \ 1991).
The observations for the interaction zone (filled triangle), for NGC~4039
(filled square) and for NGC~4038 (filled circle) are marked in the plot. 
The grey diagonal bars indicate the standard deviation of the FIR luminosity
as derived from the linear regression line of the FIR/radio correlation in
Fig.1 of Xu \etal \ (1994).
We use this scatter as an estimate for the error in distributing the total 
IRAS flux over the mapped area according to the distribution of the 20~cm
radio continuum map (see section 4.2.4).
The results for the density and the UV-intensity are given in
Table 3.
Two possible solutions for $n_{\rm H}$ and $\chi_{\rm UV}$ can be found:
a low density (l) and a high density (h) solution.
In both cases the UV-intensity is modest.
Under the condition that $\chi_{\rm UV} < 10^{2}$ (low density solution) or
$\chi_{\rm UV} / n_{\rm H} \le 10^{-3}$ cm$^{3}$ (high density solution) a 
low $I_{\rm [C II]} / I_{\rm CO}$ ratio is predicted (Wolfire, Hollenbach, 
\& Tielens 1989).
In the low $\chi_{\rm UV}$ regime the [\ion{C}{2}] emission is reduced by
low temperatures $T < 92$ K, and in the low $\chi_{\rm UV} / n_{\rm H}$
regime by the CO self-shielding which brings the [\ion{C}{2}]/CO transition
closer to the surface and therefore lowers the C$^{+}$ column density.
At the interaction zone a $^{12}$CO(2 - 1) / $^{12}$CO(1 - 0) line ratio of 
1.2 has been observed and can be interpreted as arising from gas with 
$n_{\rm H} > 10^{4}$~cm$^{-3}$ (Aalto \etal \ 1995, and Fig.~3 therein), 
supporting the high density solution in this region.
For the nuclei this ratio has not been observed.
The densities we derive for the high density solution are somewhat higher 
than what Wolfire, Tielens, \& Hollenbach (1990) deduced for Orion, M82 and 
the inner region of the Galactic Center from the PDR model.
However, we find a much lower FUV flux.
The FUV flux for the high density solution is also at the lower limit of the 
range of the FUV flux derived from the ISO-LWS observations (Fischer \etal \ 
1996).
Fischer \etal \ (1996), however, determined a density for the PDRs  
which is an order of magnitude smaller than what we derived
for the individual positions.

We derived the beam filling factor for the single-component model 
(averaged over both components).
To estimate the beam filling factor we used the ratio of the derived
$\chi_{\rm FIR}$ to the modelled $\chi_{\rm UV}$ (Table 3).
If we assume that the beam filling factor is unity and that most of the 
stellar photons are absorbed by dust grains, then $\chi_{\rm FIR} \sim 2 
\times \chi_{\rm UV}$ for B stars and $\chi_{\rm FIR} \sim \chi_{\rm UV}$ 
for O stars.
Including the beam filling factor, $\Phi $, we get therefore 
$\chi_{\rm FIR} = (1 \to 2) \times \Phi \times \chi_{\rm UV}$.
For the individual galaxies NGC 4038 and NGC 4039 where a large
fraction of late type stars exists (Bushouse \& Werner 1990, Keel \etal \
1985) the observed $\chi_{\rm FIR}$ may not only be created by UV radiation,
but also from visible light. 
Thus taking the beam filling factor as $\chi_{\rm FIR} / \chi_{\rm UV}$ gives 
an upper limit.
But there is still the uncertainty from the estimate of the FIR continuum.
From the beam filling factors we estimate source sizes of
$\approx $ 24\arcsec \ for the interaction zone which is about the total
size of the overlap region observed in CO (Stanford \etal \ 1990), and
$\approx $ 19--21\arcsec \ for the nuclei which is about twice the size
estimated from radio emission (Hummel \& van der Hulst 1986) and from CO
observations (Stanford \etal \ 1990).

From the extinction corrected hydrogen recombination lines at the
interaction zone observed with ISO (Kunze \etal \ 1996) we can also deduce 
the UV intensity originating in this region.
An equivalent single star effective temperature of the stellar ionizing 
radiation field of 44000~K has been derived from the 
[\ion{Ne}{3}]~15.6~$\mu$m~/~[\ion{Ne}{2}]~12.8$~\mu$m line ratio (Kunze 
\etal \ 1996).
This effective temperature corresponds to an O5 main sequence star.
Therefore we use the properties of such a star to derive the Lyman continuum
emission.
About 60~\% of the total stellar luminosity is emitted in the Lyman continuum
with an average photon energy of 18~eV  (Panagia 1973).
The rest is mainly emitted as far-UV radiation.
Assuming case B recombination (Osterbrock 1989) we get for the relation 
between the extinction corrected Br$\alpha$ line intensity (Kunze \etal \ 
1996) and the UV intensity 
\begin{displaymath}
I_{\rm UV} = I_{5-4} \times \frac{h \nu _{\rm UV}}{h \nu _{4-2}} \times 
\frac{j_{4-2}}{j_{5-4}} \times \frac{\sum \alpha _{n-m}}{\alpha _{4-2}}
\qquad [{\rm erg~s}^{-1}{\rm cm}^{-2}{\rm sr}^{-1}]
\end{displaymath}
where $\frac{j_{5-4}}{j_{4-2}}$ is the relative intensity of Br$\alpha$ 
to H$\beta$, 
$\sum \alpha _{n-m}$ is the total recombination coefficient, and
$\alpha _{4-2}$ the effective recombination coefficient of H$\beta$ given 
from Hummer \& Storey (1987).  
From this expression we derive a UV intensity of $I_{\rm UV} = 345 \chi 
_{\circ}$ (for standard conditions: $T = 10^{4}$~K, $n = 10^{4}$~cm$^{-3}$) 
and thus a far-UV intensity of $I_{\rm FUV} = 230 \chi_{\circ}$ ($\chi_{UV}
= 230$) which is comparable to the far-UV we obtained from the 
$Y_{\rm [C {II}]} - Y_{\rm CO}$ plot for the interaction zone.
We have done this estimate only for the interaction zone since ISO has only 
measured the hydrogen lines in the interaction zone.

\subsubsection{[C II] From Cold Neutral Medium (CNM) and/or 
Warm Neutral Medium (WNM)}

In the neutral interstellar medium ionized carbon is excited by collisions 
with electrons and atomic hydrogen.
The integrated intensity one would expect through collisions can be estimated
by (see also Madden \etal \ 1993, 1997)
\begin{displaymath}
 I_{\rm [C II]} = \frac{h \nu A}{4 \pi} \left[
 \frac{2 \exp(- \frac{91}{T})}{1 + 2 \exp(- \frac{91}{T}) + 
 \frac{n_{\rm Hcrit}}{n_{\rm H}}} +
 \frac{2 \exp(- \frac{91}{T})}{1 + 2 \exp(- \frac{91}{T}) + 
 \frac{n_{\rm ecrit}}{n_{\rm H} X_{\rm e}}} \right]
 {X_{\rm C^{+}} N_{\rm H} \Phi_{\rm b}}
\end{displaymath}
where $h$ is the Planck constant, $\nu$ is the frequency of the transition, 
$A$ is the Einstein coefficient for spontaneous emission, $2.29 \times 
10^{-6}$ s$^{-1}$ (Nussbaumer \& Storey 1981), $T$ is the temperature, 
$n_{\rm H}$ is the density of atomic hydrogen, and $X_{\rm e}$ is the 
ionization fraction of the medium.
We assume the carbon abundance to be solar
and that all
the carbon is in the form of C$^{+}$ ($X_{\rm C^{+}} \approx 3 \times
10^{-4}$).
We used 
the peak value of the column density ($N_{\rm H}$) plot of
van der Hulst (1979, Fig. 5 therein): $N_{\rm H} \approx 6.3 \times
10^{20}$ cm$^{-2}$.
The beam filling factor was assumed to be 1.
The critical density for collisions with atomic hydrogen, $n_{\rm Hcrit}$,
is deduced from the cooling function 
for collisional excitation of [\ion{C}{2}] by atomic hydrogen 
(Launay \& Roueff 1977), and that for collisions with electrons, 
$n_{\rm ecrit}$, 
is deduced from collision strengths for [\ion{C}{2}] collisions with 
electrons (Blum \& Pradhan 1992), and fitted as functions of temperature.

The warm neutral medium in the Galaxy is characterized by low density, 
$n_{\rm H} \sim 1$ cm$^{-3}$, and temperatures in the range $T \sim 4 \times 
10^{3}$ -- $8 \times 10^{3}$ K, with an ionization fraction of $X_{\rm e} 
\sim 3 \times 10^{-2}$ (Kulkarni \& Heiles 1987, 1988).
The density in this medium is far below the critical densities for
collisions of [\ion{C}{2}] with electrons and atomic hydrogen.
Thus every collisional excitation will lead to radiative deexcitation.
In this regime a maximum possible integrated [\ion{C}{2}] intensity of
$I_{\rm [C II]} \approx 1 \times 10^{-6}$ \cgs \ (at $T = 4 \times
10^{3}$~K, $n_{\rm H} = 1$ cm$^{-3}$) is estimated for collisions with both
electrons and atomic hydrogen, and both collision partners contribute
roughly equally to the integrated intensity.
This estimated value is almost two orders of magnitude lower than the
integrated intensity observed at the [\ion{C}{2}] peak which coincides
roughly with the peak of the \ion{H}{1} column density.
Since this estimated [\ion{C}{2}] integrated intensity is also more than one 
order of magnitude smaller than the observed integrated intensity in the 
nuclei, where the \ion{H}{1} column density should be smaller (van der Hulst
1979), a contribution of [\ion{C}{2}] emission from the WNM to the observed 
integrated intensity is negligible.

Carrying out the same calculations for cold neutral medium characterized 
by $T \sim 50$ -- 100~K, $n_{\rm H} \sim $ 50 -- 200~cm$^{-3}$, ionization 
fraction of $X_{\rm e} \approx 5 \times 10^{-4}$ (estimated for the Galaxy,
Kulkarni \& Heiles 1987, 1988) and again using the peak 
column density of $N_{\rm H} \approx 6.3 \times 10^{20}$~cm$^{-2}$ (van 
der Hulst 1979) we get a maximum integrated intensity of $I_{\rm [C II]} 
\approx 2.7 \times 10^{-5}$ \cgs \ (at $T = 100$~K, $n_{\rm H} = 
200$~cm$^{-3}$) and a minimum integrated intensity of $I_{\rm [C II]} 
\approx 3.0 \times 10^{-6}$ \cgs \ (at $T = 50$~K, $n_{\rm H} = 50$ 
cm$^{-3}$) for collisions with electrons and atomic hydrogen, where the 
contribution of collisions with electrons can almost be neglected. 
Therefore, at standard conditions ($T = 70$~K, $n_{\rm H} = 100$~cm$^{-3}$) 
the [\ion{C}{2}] emission from CNM is a factor of 9 smaller than the total 
[\ion{C}{2}] radiation at the interaction zone, but 
may contribute as much as $\sim \slantfrac{1}{3}$ of the total [\ion{C}{2}] 
emission.
For the nuclei we estimate a similar contribution from CNM to the 
[\ion{C}{2}] radiation using a \ion{H}{1} column density of $\approx 3.5 
\times 10^{20}$ cm$^{-2}$, from Fig.~5 of van der Hulst (1979) at the
positions of the nuclei.

\subsubsection{[C II] From Ionized Gas}

Another possibility for the origin of the [\ion{C}{2}] emission is from
ionized gas, either extended low density warm ionized medium (ELDWIM) 
or ``standard'' \ion{H}{2} regions.

Although the shape of the contour plot of the [\ion{C}{2}] fine structure
line (of total velocity range) shows some similarities to the low resolution
20~cm radio continuum (Hummel \& van der Hulst 1986) (Fig.3), 
this extended radio continuum has a relatively steep radio spectrum ($\alpha
< -0.8$) hence it is almost completely non-thermal and does not trace
extended ionized regions. 
The radio spectrum is steepest in the eastern part of the system, and 
regions devoid of strong optical emission, e.g. at the location of the 
dust patch in between the galaxy nuclei, show steeper spectra than regions 
that are bright in H$\alpha$ and blue light.
Therefore, the ELDWIM probably plays only a minor role and [\ion{C}{2}]
emission from this medium may be neglected.

However, the radio knots 1--13 seen in the 4.9~GHz high resolution map of 
Hummel \& van der Hulst (1986) (Fig. 5) have a sizeable 
thermal component.
To estimate the [\ion{C}{2}] intensity expected to arise from the thermal
emission of the knots within our beam 
we weight them relative to their distance from the center of the beam.
We calculated the emission measure (Spitzer 1978) to be $EM = 
1250$~cm$^{-6}$pc for the
interaction zone and $EM = 270$~cm$^{-6}$pc and $EM = 690$~cm$^{-6}$pc for 
NGC~4038 and NGC~4039, respectively.
Making the crude approximation that all of the carbon in the \ion{H}{2}
region is in form of C$^{+}$ and that all of the observed $EM$ results in 
the excitation of [\ion{C}{2}], the expected [\ion{C}{2}] integrated 
intensity is
\begin{displaymath}
 I_{\rm [C II]} = \frac{h \nu A}{4 \pi n_{\rm crit}} \left[
 \frac{\frac{g_{u}}{g_{l}}}{1 + \left( 1 + \frac{g_{u}}{g_{l}} \right)
 \frac{n_{\rm e}}{n_{\rm crit}}} \right] X_{\rm C^{+}} EM \qquad .
\end{displaymath}
Here $g_{u}/g_{l} = 2$ is the ratio of the statistical weights in the upper 
and lower level and $n_{\rm crit} = 49$ cm$^{-3}$ is the critical density of 
C$^{+}$ for collisions with electrons at $10^{4}$ K (Blum \& Pradhan 1992). 
Assuming an electron density of $n_{\rm e} = 300$ cm$^{-3}$ as estimated 
from the extinction corrected ratio of the [\ion{S}{3}] 
18.71$\mu$m/33.48$\mu$m fine structure line (Kunze \etal \ 1996) 
the expected [\ion{C}{2}] intensity is
$\sim 5.6 \times 10^{-6}$ \cgs \ at the interaction zone, which is about 
a factor 20 smaller than the observed [\ion{C}{2}] intensity in the lower 
velocity range. 
For the nuclei we assume an electron density of $n_{\rm e} = 100$ cm$^{-3}$ 
(Rubin \etal \ 1970) and obtain expected [\ion{C}{2}] intensities of $I_{\rm 
[C II]} = 3.3 \times 10^{-6}$ \cgs \ and $I_{\rm [C II]} = 8.4 \times 
10^{-6}$ \cgs \ for NGC 4038 and NGC 4039, respectively.
This is about a factor 17 lower for NGC 4038 and a factor 8 lower for NGC
4039 than the observed [\ion{C}{2}] intensities at the upper velocity range.
However, most of the carbon in \ion{H}{2} regions is probably in the form of 
C$^{++}$ rather than C$^{+}$. 
Therefore, the contribution of 
the [\ion{C}{2}] emission expected from \ion{H}{2} regions is even lower
and should be only a minor fraction of the observed [\ion{C}{2}] 
radiation.

\section{Discussion}

As shown in the previous section a possible
contribution of [\ion{C}{2}] emission from the WNM, ELDWIM and \ion{H}{2} 
regions to the total observed [\ion{C}{2}] emission is small.
For high temperatures and high densities a contribution of [\ion{C}{2}] 
emission from CNM may become important.
However, since at standard conditions of the CNM a contribution of 
[\ion{C}{2}] emission to the observed [\ion{C}{2}] emission is also small
we will only consider PDRs as the origin of the [\ion{C}{2}] emission in the
following discussion.

Stacey \etal \ (1991) demonstrated the utility of the [\ion{C}{2}]/CO(1-0)
ratio to distinguish between starburst activity in galaxies and more 
quiesent regions.
They find [\ion{C}{2}]/CO $\sim 6100$ 
(4000)\footnote[1]{CO data (NRAO) corrected for main beam 
efficiency ($\sim 0.65$)} for starburst nuclei and Galactic 
starforming regions while the more quiesent regions show a ratio of 
$\sim 2000$ (1300)\footnotemark[1].
We find [\ion{C}{2}]/CO $\sim 2350$ (1650)\footnote[2]{CO data 
(SEST) corrected for main beam efficiency ($\sim 0.7$)} toward the 
[\ion{C}{2}] peak (overlap region) and 1650 to 1800 
(1150 to 1250)\footnotemark[2] toward the nuclei, implying 
there is no strong starburst activity taking place, given the resolution 
of our data.
These ratios are more consistent with quiescent spiral galaxies, e.g. 
NGC~891 and NGC~3628 (Stacey \etal \ 1991).
However, observations with ISO (Kunze \etal \ 1996; Fischer \etal \ 1996;
Vigroux \etal \ 1996), CO observations (Stanford \etal \ 1990; Aalto
\etal \ 1995), and observations of the radio continuum (Hummel \& van der
Hulst 1986) indicate ongoing strong star formation activity in the 
interaction zone.
A comparison of the CO flux of the interferometric observation
(scaled to the size and the position of the single dish observation) of
Stanford \etal \ (1990) (Fig. 2) with the single dish 
observation of Aalto \etal \ (1995) show a factor of about 4 more CO flux in 
the single dish observation ($F_{\rm CO}^{O}$) for the interaction zone and 
NGC~4039 and a factor 3 for NGC~4038.
If we assume that $\slantfrac{1}{4}$ of the single dish CO flux at the 
interaction zone arises from 
a confined starburst region, $F_{\rm CO}^{\rm SB} = F_{\rm CO}^{O} 
\times \slantfrac{1}{4}$, and use a ratio of 
$F_{\rm [C II]}^{\rm SB}/F_{\rm CO}^{\rm SB} = 6000$ (the mean ratio 
observed for starburst galaxies, Stacey \etal \ 1991) for this region 
we find a ratio of $(F_{\rm [C II]}^{\rm SB} - 
F_{\rm [C II]}^{O})/(F_{\rm CO}^{\rm SB} - F_{\rm CO}^{O})
\approx 1200$ for an underlying component.
For $F_{\rm [C II]}^{O}/F_{\rm CO}^{O}$ we used the observed
ratio of 2350.
This estimate suggests that the [\ion{C}{2}] emission toward the interaction
zone can also be explained
by a two-component model with a confined starburst region and a quiescent
surrounding molecular cloud system.
Making the same estimate for NGC~4039 we get a [\ion{C}{2}] to CO ratio of 
400 for an underlying component.
To make the same estimate for NGC~4038 one has to assume a lower 
[\ion{C}{2}] to CO integrated intensity ratio for the confined starburst
region since otherwise the ratio of the underlying component becomes 
negative.
The lowest value for the [\ion{C}{2}] to CO ratio in Stacey \etal \ (1991)
is 340 for the SgrA +20~km~s$^{-1}$ cloud.
If we assume a lower limit for the [\ion{C}{2}] to CO ratio for the 
underlying component of 300 we find a maximum [\ion{C}{2}] to CO ratio for 
a confined starburst region of 4400 for NGC~4038 
using the observed ratio of $F_{\rm [C II]}^{O}/F_{\rm CO}^{O} = 1650$.
From this estimate we find that an enhanced star forming activity in NGC~4039
and a moderate activity in NGC~4038 could be explained by the data.
This interpretation would be consistent with the results of the mid-infrared
observations of NGC 4038/39 with ISOCAM (Vigroux \etal \ 1996). 
They show that the most active star formation in the Antennae system occurs 
in the overlap region, and within that region in a confined area (knot 2 in
Fig. 5; knot A in 
Fig. 1 in Vigroux \etal \ 1996) coinciding with the southern clump of the CO 
emission in Stanford \etal \ (1990).
The ISO SWS observations of the interaction zone can also be described by a 
young ($\approx 7 \times 10^{6}$~yr) starburst with an IMF extending up to 
100~M$_{\sun}$ (Kunze \etal \ 1996).
The interpretation that the [\ion{C}{2}] emission arises from confined small
star forming regions is also consistent with the detection of Br$\gamma$ knots
in the interaction zone (Fischer \etal \ 1996).
The case that most of the [\ion{C}{2}] emission comes from confined star forming
regions supports the PDR solution with a small beam filling factor and 
therefore with high density.
The PDR solution with high density is also supported by the ISO observations of
Fischer \etal \ (1996).

Although we see moderate [\ion{C}{2}] emission from the \ion{H}{2} regions 
in the western loop around NGC~4038, in the absence of CO data for that 
region and taking our beam size into account we cannot draw a conclusion 
about the star forming activity there.

\section{Conclusion}

We present a map of NGC 4038/39 in the [\ion{C}{2}] 158 \micron \ fine 
structure line. 
[\ion{C}{2}] emission is detected over the optical extent of the system of
galaxies and peaks at the interaction zone.
The total luminosity of the [\ion{C}{2}] line is $3.7 \times 10^{8} 
L_{\sun}$ which is about 1\% of the FIR luminosity of the Antennae. 
Only a 
negligible fraction of the observed [\ion{C}{2}] emission can originate in 
the WNM, if conditions are similar to Galactic atomic clouds.
Under normal conditions the [\ion{C}{2}] emission from standard CNM makes 
only a small contribution to the total [\ion{C}{2}] emission, 
however it may rise to $\slantfrac{1}{3}$
of the total [\ion{C}{2}] emission for individual positions. 
Only minor fractions of the [\ion{C}{2}] emission at the interaction zone 
and the nuclei can arise from \ion{H}{2} regions.
PDRs are the dominant source for the [\ion{C}{2}] emission.
We estimate minimum hydrogen masses associated with the [\ion{C}{2}] 
emitting region of $1.9 \times 10^{8} M_{\sun}$ for the entire merging 
system and $6.8 \times 10^{7} M_{\sun}$, $3.2 \times 10^{7} M_{\sun}$, and 
$3.7 \times 10^{7} M_{\sun}$ within one beam centered at the interaction 
zone, NGC~4038, and NGC~4039, respectively. 
Assuming a single emission component in the beam we derive a density of the 
[\ion{C}{2}] emitting gas in PDRs of $1 \times 10^{5}$ cm$^{-3}$ for the 
interaction zone and of $2 \times 10^{5}$ cm$^{-3}$ and $1 \times 
10^{5}$ cm$^{-3}$ for NGC~4038 and NGC 4039, respectively, and a 
far-UV intensity of $450 \chi_{\circ}$ for the interaction zone, 
$500 \chi_{\circ}$ for NGC~4038, and $250 \chi_{\circ}$ for NGC~4039. 
The derived beam filling factor of the emission from the PDRs is 20\% for 
the interaction zone and 10--15\% for the nuclei.
However, the PDR model also allows a solution with low density 
($\sim 10^{2}$~cm$^{-3}$), high beam filling factor ($\sim 50$\%), and low
FUV intensity ($\sim 100 \chi_{\circ}$).
The low, beam averaged [\ion{C}{2}]/CO ratio of 2350 toward the interaction 
zone and the even lower ratios at the nuclei indicate that no global 
starburst is going on either in the area surrounding the interaction zone or 
in the nuclear region. 
The high-excitation lines observed with ISO SWS which trace the starburst 
must therefore arise from a small, confined region in the interaction zone.
This result is also supported from the observations with ISOCAM.
Therefore on the scale of our [\ion{C}{2}] beam a single emission component 
for the PDR is only a crude approximation. 
Using interferometric and single dish CO observations and an expected 
[\ion{C}{2}]/CO ratio for starburst regions and for quiescent clouds, we
constructed a two-component model consisting of a confined starburst region
and of molecular clouds enveloping the starburst.
From this model we find that the [\ion{C}{2}] emission at the interaction 
zone originates partly from confined starburst regions and partly from 
surrounding quiescent clouds.
If we apply this model to the nuclei we get also enhanced star formation
activity in NGC~4039 with a low [\ion{C}{2}]/CO ratio for the quiescent clouds
but only moderate star forming activity in NGC~4038.
This two-component model supports the high-density solution for the PDRs.

Future investigation of the Antennae in the [\ion{N}{2}] 205 \micron \ fine 
structure line would be very helpful to further disentangle the origins of 
the [\ion{C}{2}] line.
Also observations at higher spatial resolution in the FIR regime (e.g. with
FIRST and SOFIA) would be a great step forward to investigate this and other
spatially very complex objects in more detail.

\acknowledgments                                     
We are grateful to the staff of the KAO for their competent support. 
This work was partially supported by the NASA grant NAG-2-208 to the 
University of California, Berkeley. N.G. was supported in part by a
Feodor-Lynen-fellowship of the Alexander von Humboldt foundation.

\clearpage

\begin{deluxetable}{lccccc}
\scriptsize
\tablewidth{0pc}
\tablenum{1}
\tablecaption{Integrated intensity in [C{\rm II}] compared with 
CO ($1\to 0$) data \label{tab:c2res}}
\tablehead{
\colhead{Source} & \colhead{$I_{\rm [C II]}$} & \colhead{{$I_{\rm [C II]}$} 
\tablenotemark{a}} & \colhead{{$I_{\rm [C II]}$} \tablenotemark{a}} & 
\colhead{{$I_{\rm CO}$} \tablenotemark{b}}
& \colhead{$I_{\rm [C II]}/I_{\rm CO}$ \tablenotemark{c}} \nl
\colhead{} & \colhead{[$10^{-5}$ erg s$^{-1}$} 
& \colhead{[$10^{-5}$ erg s$^{-1}$} & \colhead{[$10^{-5}$ erg s$^{-1}$}
& \colhead{[$10^{-8}$ erg s$^{-1}$} & \colhead{} \nl
\colhead{} & \colhead{cm$^{-2}$sr$^{-1}$]} & \colhead{cm$^{-2}$sr$^{-1}$]}
& \colhead{cm$^{-2}$sr$^{-1}$]} & \colhead{cm$^{-2}$sr$^{-1}$]} 
& \colhead{} \nl
\colhead{} & \colhead{(1420--1740~km~s$^{-1}$)} 
& \colhead{(1550--1740~km~s$^{-1}$)} & \colhead{(1420--1554~km~s$^{-1}$)}
& \colhead{} & \colhead{}
 }
\startdata
[C {\rm II}] peak & 11.7 & 6.8 & 3.8 & 4.9$^{d}$ & 2350 \nl
(Interaction zone) & & & & (7.0)$^{e}$ & (1650)$^{e}$ \nl
NGC 4038 & 5.5 & 5.0 & 1.3 & 3.4 & 1650 \nl
 & & & & (4.8)$^{e}$ & (1150)$^{e}$ \nl
NGC 4039 & 6.3 & 3.9 & 2.6 & 3.5 & 1800 \nl
 & & & & (5.0)$^{e}$ & (1250)$^{e}$ \nl
\tablenotetext{a}{Note that there is some overlap between the two subdivided 
velocity ranges due to the limited spectral resolution.} 
\tablenotetext{b}{CO ($1\to 0$) measurements of Aalto \etal \ (1995), scaled 
to the same units as [C {\rm II}] intensity: \\
 \hspace*{1em}1~K~km~s$^{-1}~=~1.6~\times~10^{-9}$~\cgs.}
\tablenotetext{c}{$I_{\rm [C II]}$ of the total velocity range 
(1420--1740~km~s$^{-1}$)}
\tablenotetext{d}{the position of the beam of the interaction zone of
Aalto \etal \ (1995) coincides with the [C{\rm II}] peak in the total \\
 \hspace*{1em}velocity range}
\tablenotetext{e}{CO data corrected for main beam efficiency (0.7)}
\enddata
\end{deluxetable}

\vfill

\clearpage

\begin{deluxetable}{lcc}
\tablewidth{0pc}
\tablenum{2}
\tablecaption{Minimum hydrogen mass and column density \label{tab:min}}
\tablehead{
\colhead{Source} & \colhead{$M_{\rm min}({\rm H})$} &
\colhead{$N_{\rm min}({\rm H})$} \nl
\colhead{} & \colhead{$10^{7} M_{\sun}$} & \colhead{$10^{20}$ cm$^{-2}$}
 }
\startdata
[C {\rm II}] peak & 6.8 & 2.3 \nl
(Interaction zone) & & \nl
NGC 4038 & 3.2 & 1.1 \nl
NGC 4039 & 3.7 & 1.3 \nl
total galaxy pair & 18.8 & 0.67 (average) \nl
\enddata
\end{deluxetable}

\vfill

\clearpage

\begin{deluxetable}{lcccc}
\tablewidth{0pc}
\tablenum{3}
\tablecaption{PDR model output \label{tab:pdrout}}
\tablehead{
\colhead{Source} & \colhead{PDR-} & 
\colhead{$n_{\rm H}$} &
\colhead{$\chi_{\rm FUV}$ \tablenotemark{b}} & 
\colhead{$\Phi_{\rm b}$ \tablenotemark{c}} \nl
\colhead{} & \colhead{solution \tablenotemark{a}} & \colhead{cm$^{-3}$} &
\colhead{} & \colhead{}
 }
\startdata
Interaction zone & l & $4\times 10^{2}$ & 150 & 0.6 \nl
([C{\rm II}]) peak & & & & \nl
                 & h & $1\times 10^{5}$ & 460 & 0.2 \nl
NGC 4038 & l & $1.4\times 10^{2}$ & 120 & 0.50 \nl
       & h & $2\times 10^{5}$ & 500 & 0.1 \nl
NGC 4039 & l & $3\times 10^{2}$ & 80 & 0.45 \nl
        & h & $1\times 10^{5}$ & 240 & 0.15 \nl
\tablenotetext{a}{l: low density solution \\
 \hspace*{1em}h: high density solution}
\tablenotetext{b}{In units of $\chi_{\circ} = 2\times 10^{-4}$ erg s$^{-1}$ 
cm$^{-2}$ sr$^{-1}$ \\
 \hspace*{1em}(Draine 1978)}
\tablenotetext{c}{Beam filling factor: $\Phi_{\rm b} = \frac{\chi_{\rm 
FIR}}{\chi_{\rm FUV}}$}
\enddata
\end{deluxetable}

\vfill

\clearpage

\begin{figure}
 \figurenum{1}
 \label{fig:C IIop1}
 \caption[]{
Integrated [C II]~158 \micron \/ intensity map in the velocity ranges of 
a)~1420--1740~km~s$^{-1}$, the total [\ion{C}{2}] emission, 
b)~1550--1740~km~s$^{-1}$, the velocity range containing the 2 nuclei, and
c)~1420--1554~km~s$^{-1}$, the velocity range of the interaction zone,
superimposed on an optical image of NGC 4038/39 (Laustsen, Madsen, \& 
West 1987). 
The contour levels in all three maps are the same, starting at 1 $\sigma$ 
with steps of 1 $\sigma$ ($1.04 \times 10^{-5}$ 
erg~s$^{-1}$cm$^{-2}$sr$^{-1}$).
The hatched circle indicates the beam (FWHM of 55\arcsec).
The symbol ``S'' indicates the southern clump of the interaction zone in 
Stanford \etal \ (1990).}
\end{figure}

\begin{figure}
 \figurenum{2}
 \label{fig:CCO}
 \caption[]{
CO (1-0) map of Stanford \etal \ (1990)
(white contours) superimposed on 
the [C II] integrated 
intensity map of the total velocity range (black contours).
The FWHM of the [C II] beam, indicated by the hatched circle, is 55\arcsec.
S, W, N, E, indicate the southern, western, northern, and eastern CO clump 
at the interaction zone (Stanford \etal \ 1990).
The grey area represents the area covered by the [\ion{C}{2}] arrays.
(Note: The levels of the CO contours at NGC~4038 are different from the 
levels at the interaction zone and NGC~4039)}
\end{figure}

\begin{figure}
 \figurenum{3}
 \label{fig:C II20}
 \caption[]{
1.5~GHz continuum map of Hummel \& van der Hulst (1986) 
(white contours) superimposed on 
the [C II] integrated 
intensity map of the total velocity range (black contours).
The FWHM of the [C II] beam, indicated by the hatched circle, is 55\arcsec;
the HPBW of the radio map is 16\arcsec.
The grey area represents the area covered by the [\ion{C}{2}] arrays.}
\end{figure}

\begin{figure}
 \figurenum{4}
 \label{fig:pdr}
 \caption[]{
Ratios $Y_{\rm [C II]} = I_{\rm [C II]} / \chi_{\rm FIR}$ vs.\ 
$Y_{\rm CO} = I_{\rm CO} / \chi_{\rm FIR}$.
The solid lines represent a set of constant densities and the dashed lines a 
set of constant UV intensities characteristic of PDRs (Stacey 1991).
The filled triangle marks the observations at the [C II] peak, the filled
square NGC~4039, and the filled circle NGC~4038.
The grey diagonal bars indicate the standard deviation of $\chi_{\rm FIR}$.}
\end{figure}

\begin{figure}
 \figurenum{5}
 \label{fig:C II6}
 \caption[]{
4.9~GHz continuum map of Hummel \& van der Hulst (1986) 
(white contours) superimposed on 
the [C II] integrated 
intensity map of the total velocity range (black contours).
The FWHM of the [C II] beam, indicated by the hatched circle, is 55\arcsec;
the HPBW of the radio map is 6\arcsec.
The numbers 1~--~13 indicate the discrete radio knots (Hummel \& van der 
Hulst 1986).
The grey area represents the area covered by the arrays.}
\end{figure}

\vfill

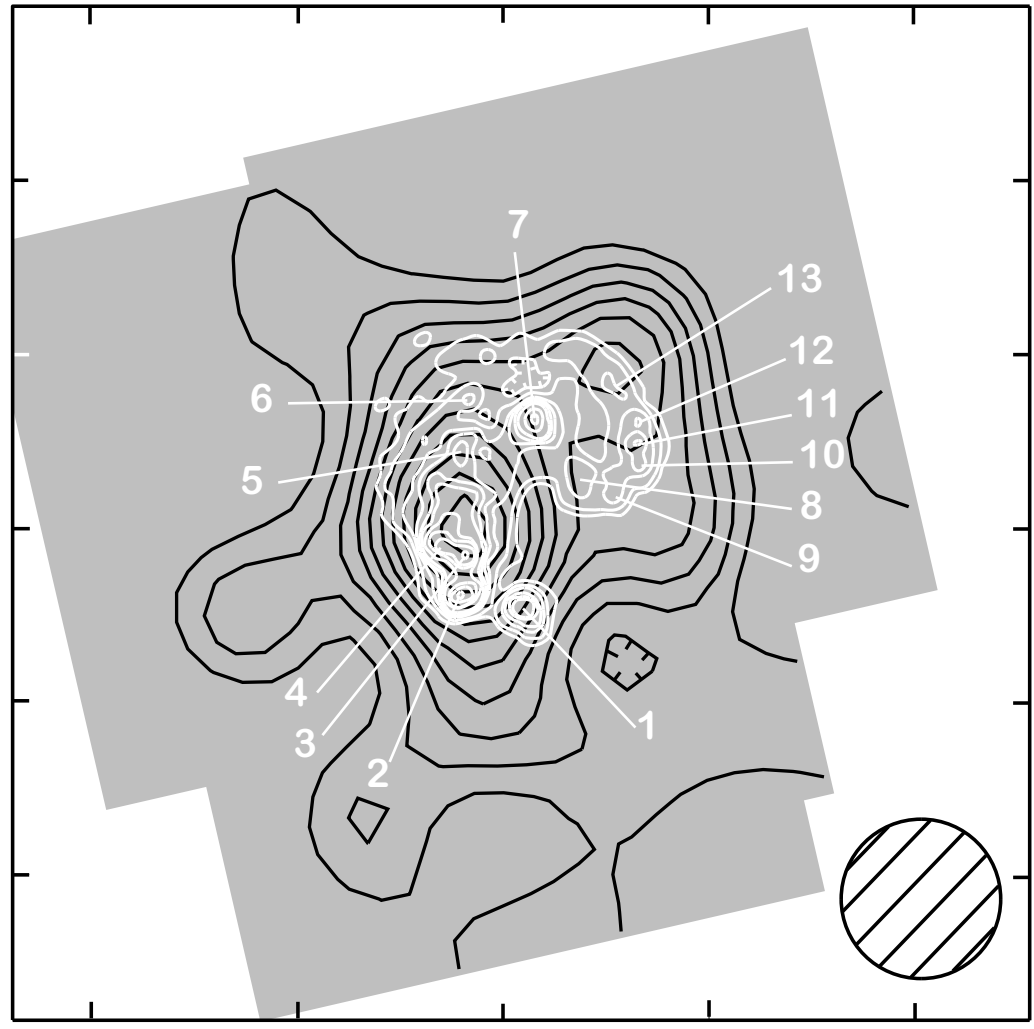
\end{document}